\documentclass[aps, prl, 
superscriptaddress,
nobibnotes,
amsmath,
amssymb,
reprint]{revtex4-2}
\usepackage{graphicx}
\usepackage{ninecolors} 
\usepackage{xcolor, hyperref} 
\usepackage{bm, physics}
\usepackage{subfiles}  
\usepackage{mathtools}
\usepackage{graphicx}
\newcommand{\upstate}{\ensuremath{\ket{\uparrow}}}
\newcommand{\downstate}{\ensuremath{\ket{\downarrow}}}

\begin{document}
\preprint{APS/123-QED}
\title{High-fidelity entanglement of polar molecules by dynamic geometric control}

\author{Scarlett S. Yu}
\affiliation{Department of Physics, Harvard University, Cambridge, MA 02138, USA}
\affiliation{Harvard-MIT Center for Ultracold Atoms, Cambridge, MA 02138, USA}

\author{Avikar Periwal}
\affiliation{Department of Physics, Harvard University, Cambridge, MA 02138, USA}
\affiliation{Harvard-MIT Center for Ultracold Atoms, Cambridge, MA 02138, USA}
\affiliation{Department of Physics, Massachusetts Institute of Technology, Cambridge, MA 02139, USA}

\author{Jiaqi You}
\affiliation{Department of Physics, Harvard University, Cambridge, MA 02138, USA}
\affiliation{Harvard-MIT Center for Ultracold Atoms, Cambridge, MA 02138, USA}

\author{Zirui Liu}
\affiliation{Institute for Interdisciplinary Information Sciences, Tsinghua University, Beijing 100084, PR China}

\author{Qinshu Lyu}
\affiliation{Centre for Cold Matter, Blackett Laboratory, Imperial College London, London SW7 2AZ, United Kingdom}

\author{Youngju Cho}
\affiliation{Department of Physics, Korea University, Seongbuk-gu, Seoul 02841, South Korea}

\author{Lo\"ic Anderegg}
\affiliation{Department of Physics and Astronomy, University of Southern California, Los Angeles, CA 90089, USA}

\author{Eunmi Chae}
\affiliation{Department of Physics, Korea University, Seongbuk-gu, Seoul 02841, South Korea}


\author{John M. Doyle} 
\affiliation{Department of Physics, Harvard University, Cambridge, MA 02138, USA}
\affiliation{Harvard-MIT Center for Ultracold Atoms, Cambridge, MA 02138, USA}

\begin{abstract}
In quantum information systems made of optical tweezer arrays of ultracold molecules, thermal motion of molecules degrades the coherence of their interactions, which limits entanglement fidelity and the concomitant scientific applicability of these systems. We show that by controlling the geometry of the dipolar interaction, even when a molecule occupies many motional states in the tweezer, coherence can be preserved. We characterize several geometries that suppress sensitivity to thermal fluctuations. We further use programmable, coherence-preserving motion of the molecules during entanglement to refocus dephasing from relative positional jitter of the tweezers, which is relevant even on the 10\,nm scale. These methods yield substantially improved dipolar coherence and enable generation of two-molecule entanglement with a Bell state fidelity of $\mathcal{F}= 0.976^{+0.008}_{-0.011}$ in directly laser-cooled molecules.
\end{abstract}

\maketitle

\section{Introduction}
A defining requirement for quantum simulation and quantum information processing is the ability to generate entanglement through controlled interactions. Polar molecules offer a compelling combination of quantum resources, including long-lived rotational transitions~\cite{park2017second, burchesky2021rotational, park2023extended, gregory2024second, hepworth2025long}, rich internal state manifolds for multilevel encoding~\cite{sawant2020qudits, sundar2018synthetic, hepworth2025long}, and molecular electric dipole moments that enable long-range entangling interactions~\cite{bao2023dipolar, holland2023demand, picard2025entanglement, ruttley2025long}. Notably, rotational qubits can be easily tuned and controlled using microwave and dc electric fields~\cite{li2021tuning, li2023tunable, miller2024two, lu2026probing}. The capabilities of these qubits suggest new routes to quantum simulation of many-body spin dynamics~\cite{micheli2006toolbox,gorshkov2011tunable,yan2013observation,hazzard2014many, gadway2016strongly,christakis2023probing, carroll2025observation, lu2026probing, bilitewski2021dynamical}, quantum information processing~\cite{demille2002quantum,yelin2006dipolarQC,karra2016paramagnetic,ni2018dipolar, sawant2020qudits,cornish2024quantum}, and precision laboratory searches for both dark matter~\cite{kozyryev2021enhanced, demille2024quantum} and new CP-violating particles in the $>10$\,TeV range~\cite{chupp2019electric, cairncross2019atoms, kozyryev2017precision, hutzler2020polyatomic}. Recent experiments have realized dipolar spin exchange interactions and Bell state creation between individual molecules in optical tweezers~\cite{bao2023dipolar, holland2023demand, christakis2023probing, picard2025entanglement, ruttley2025long}, placing molecular systems in a practical regime for quantum simulation of interacting spin models and entangling operations for digital quantum information processing.

The coherent quantum dynamics of two interacting molecules is driven by the dipolar coupling ($J$), with strength that depends on the relative displacement vector $\vec{r}$ between the molecules and their orientation with respect to the quantization field axis. Changes in $\vec{r}$ translate into changes in the entanglement rate. A central limitation for coherent dipolar driven entanglement is dephasing induced by thermal motion of molecules. Thermal occupation of many motional states leads to a distribution of effective interaction strengths. For molecules trapped in optical tweezers, the weak confinement along the longer tweezer beam axis makes the molecular positional wave packet strongly anisotropic, and is the dominant contribution to dephasing for previous demonstrations of entanglement in polar molecules~\cite{picard2025entanglement, ruttley2025long}. The few quanta of motional excitation provided a fundamental limit on interaction coherence. This motivates finding new strategies to suppress sensitivity of $J$ to thermal fluctuations by narrowing the distribution of effective coupling strengths sampled across experimental realizations to improve many-body coherence~\cite{ni2018dipolar, you2025control, emperauger2025benchmarking,bergonzoni2025iswap}.
\begin{figure*}[t!] 
    \centering
    \includegraphics[width=1\textwidth]{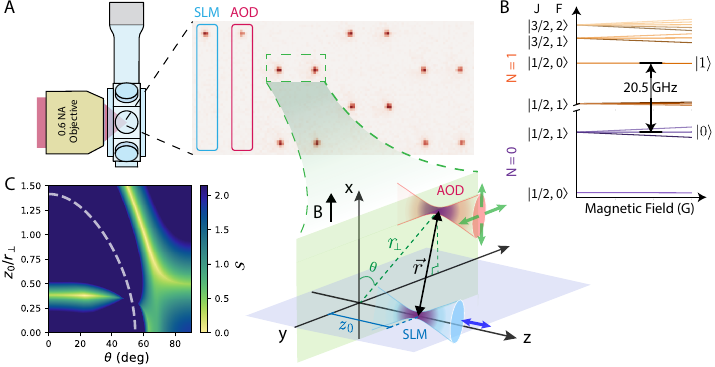}
        \caption{\textbf{Three-dimensional geometric control of dipolar interactions in a molecular tweezer array.} (\textbf{A}) Schematic of the optical tweezer array and geometric coordinates used in this work. CaF (calcium monofluoride) molecules are loaded into 775\,nm optical tweezers projected through a 0.6\,NA microscope objective into a vacuum glass cell. Inset: Geometry and coordinates between one pair of interacting molecules. The tweezer propagation is along $\hat{z}$. The relative separation vector between two molecules is $\vec{r} = \vec{r}_\perp + z_0 \hat{z}$, where $\vec{r}_\perp$ is the in-plane separation and $z_0$ is their relative axial displacement. The angle $\theta$ is defined between $\vec{r}_\perp$ and the quantization axis, which is set by a small applied magnetic field along $\hat{x}$. One trap (shown in red) is generated from crossed 2D-AODs, which provide lateral positioning to set $(r_{\perp}, \theta)$, while the other trap (shown in blue) is projected from an SLM phase mask, which can independently impose $z_0$. (\textbf{B}) Relevant internal state manifolds used in preparing and defining the rotational qubit. (\textbf{C}) Dipolar motional sensitivity in experimentally accessible three-dimensional geometry. Color shows the dimensionless weighted relative sensitivity $\mathcal{S}$, defined in main text, of the interaction strength to thermally-induced dephasing. The dashed curve marks $J= 0$; brighter regimes indicate reduced leading-order sensitivity to thermal motion. }
        \label{fig:fig1}
\end{figure*}
In this work, we utilize both static and dynamic geometrical programming to control and improve coherence in two-qubit entanglement of molecules in an optical tweezer array. We first experimentally demonstrate that static geometry engineering can suppress thermal dephasing by realizing interaction configurations that minimize sensitivity to motional fluctuations. Under optimal conditions, trap position fluctuations on the order of a few nanometers become the dominant source of dephasing. We take advantage of the long rotational coherence time of molecules to implement a geometric echo protocol that changes the relative molecular positions during dipolar entangling, effectively refocusing positional noise. This demonstration of coherent molecular locomotion highlights that programmable trap motion can be used not only for assembly and rearrangement, but as a coherent rectifying operation during interactions. With the addition of partial Raman sideband cooling, higher-order thermal effects are suppressed to realize a two-molecule Bell state with a measured fidelity of $\mathcal{F}=0.976^{+0.008}_{-0.011}$. This result, to our knowledge, is the highest reported fidelity for directly laser-cooled molecules, and is comparable to state-of-the-art demonstrations of entanglement between assembled molecules~\cite{picard2025entanglement, ruttley2025long}. The power of these methods is highlighted because the molecular temperatures here are much higher. This sets a firm, realistic path to fidelities well above the 99\% level, as first envisioned by theory~\cite{ni2018dipolar, you2025control}.

Our experiment is carried out with laser-cooled calcium monofluoride (CaF) molecules individually trapped in a two-dimensional (2D) optical tweezer array, illustrated in Fig.~\ref{fig:fig1}A. We first describe the tweezer architecture relevant to geometric control of dipolar interactions. (Molecule production, laser cooling, and trapping of CaF have been described in detail in previous work~\cite{bao2022fast, yu2024conveyor}). The tweezer array is formed from two beam paths derived from the same laser: in one path, the light is reflected by a spatial light modulator (SLM), which generates a static array, while the other beam is diffracted by two crossed acousto-optic deflectors (2D-AODs) to create a set of dynamic tweezers. The two beams are combined and projected through the microscope objective. One trap of an interacting pair of molecules is generated by the SLM and the other by the AODs. The SLM defines a static two-dimensional array and the trap position in the axial direction, while the AODs provide rapid motion in the lateral directions. This architecture allows convenient access to the three-dimensional interaction geometry between pairs of directly laser-cooled CaF molecules, and avoids using multi-tone AOD operation which can introduce parametric heating~\cite{endres2016atom}. 

Molecules are initially loaded and rearranged into a 2D array in the SLM tweezers with 14\,$\mu$m spacing. We encode a qubit in two rotational states $\upstate \equiv \lvert {X; \nu = 0, N = 1, F = 0, m_{F} = 0} \rangle$ and $\downstate \equiv \lvert {X; \nu = 0, N = 0, F = 1, m_{F} = 0} \rangle$ within the electronic and vibrational ground state $X^2\Sigma^{+}$ of CaF (see Fig.~\ref{fig:fig1}B). All molecules are initialized in the qubit state $\upstate$. We then transfer half of the molecules from the SLM tweezers into the dynamic AOD tweezers (see Fig.~\ref{fig:fig1}A), and move each molecule toward its neighboring SLM molecule, forming pairs of interacting molecules. A magnetic field of $\sim$ 3\,G applied along $\hat{x}$ defines the quantization axis. Each pair of molecules interacts with a dipolar coupling strength of 
\begin{equation}
    J = \frac{d^2}{4\pi\epsilon_0 r^3} (1-3 \cos^2\alpha)
\end{equation}
where $d$ is the transition dipole moment for the $\downstate \leftrightarrow \upstate$ transition, $r = |\vec{r}|$ is the intermolecular separation, and  $\alpha$ is the angle between $\vec{r}$ and the quantization axis. In most implementations, this coupling is tuned by changing $r$ and, when angular control of $\alpha$ is needed, by reorienting the pair either in the transverse plane or by rotating the quantization field itself. In our apparatus, the three-dimensional degrees of freedom of an individual pair are specified through experimentally accessible parameters in cylindrical coordinates $(r_{\perp}, \theta,  z_0)$, where  $r_{\perp}$ is the magnitude of the projection of $\vec{r}$ onto the transverse $x-y$ plane, $\theta$ is its azimuthal angle relative to $\hat{x}$, and $ z_0$ is the relative axial displacement (Fig.~\ref{fig:fig1}A, inset). The 2D-AODs provide rapid lateral positioning to set $(r_{\perp}, \theta)$, and the SLM hologram can independently impose $z_0$.

With the interaction geometry set, a Ramsey-$\pi/2$ microwave pulse is applied to prepare each pair of molecules in a superposition $(\upstate+\downstate)/\sqrt{2}$, which evolves under the dipolar Hamiltonian for a variable time. During this time we apply an XY8 dynamic decoupling sequence~\cite{bao2023dipolar} to remove sources of single-body decoherence while preserving the dipolar interaction. A final Ramsey-$\pi/2$ pulse maps the dynamics onto two-qubit populations $(P_{\downarrow\downarrow}, P_{\uparrow\uparrow}, P_{\downarrow\uparrow}, P_{\uparrow\downarrow})$, corresponding to the probabilities of finding the pair in ($\lvert{\downarrow\downarrow}\rangle, \lvert{\uparrow\uparrow}\rangle, \lvert{\downarrow\uparrow}\rangle, \lvert{\uparrow\downarrow}\rangle$), respectively.

\subsection*{Static geometric engineering of motional sensitivity}
We begin by studying the motional sensitivity of the dipolar interaction and its dependence on geometry. Thermal motion of molecules broadens the distribution of the dipolar interaction $J$ across the thermal ensemble by modifying both the intermolecular separation $r$ and the angle $\theta$. For CaF molecules in optical tweezers, the motional timescales set by the trap frequencies are orders of magnitude larger than the dipolar exchange dynamics. In this regime, the motional distribution is symmetric about the trap center and the linear sensitivity $\nabla J \cdot \delta \vec{r}$ averages to zero. The leading contribution to $J$ is set by the width of the molecular wave packet, which is sampled from a thermal distribution~\cite{you2025control}. Engineering the geometry of the interacting molecules is aimed at minimizing the quadratic sensitivity to the spatial extent of the molecular wave packet, while further cooling reduces the extent itself. Shifts of the mean trap positions, by contrast, enter through the linear term and are a first-order source of interaction dephasing.

In the simple planar configuration we first implement here - the two trap centers of an interacting molecule pair are constrained to the same focal plane ($z_0 = 0$) - the interaction geometry is set by a single in-plane angle $\theta$, which can be tuned to the planar ``magic" angle $\theta_m \approx 63.4^\circ$, at which leading-order sensitivity of $J$ to axial thermal motion vanishes, while the mean dipolar interaction remains nonzero~\cite{you2025control}. This planar magic-angle condition, however, is only one slice of a more general three-dimensional geometry dependence and cannot additionally suppress dephasing arising from radial thermal motion. Introducing a finite axial displacement $\Delta z$ between the two molecules adds an additional degree of freedom, allowing sensitivities of $J$ to thermal motion along different spatial directions to be tuned simultaneously. To characterize this dependence, we define a dimensionless weighted relative sensitivity $\mathcal{S} \equiv \sum_{i}w_i|A_{i,2}|$, where $A_{i,2} \equiv (r^2/2J)\partial^2_{i}J$ quantifies the leading-order (quadratic) change in $J$ to displacement along direction $i$, and the normalized weights $w_i \propto 1/\omega_{i}^2$ account for trap anisotropy by weighting each axis by its relative thermal extent. In Fig.~\ref{fig:fig1}C we show $\mathcal{S}(\theta, z_0)$, where brighter regions correspond to geometries with reduced overall sensitivity of $J$ to thermal motion. 
\begin{figure}[ht] 
\includegraphics[width=1\columnwidth]{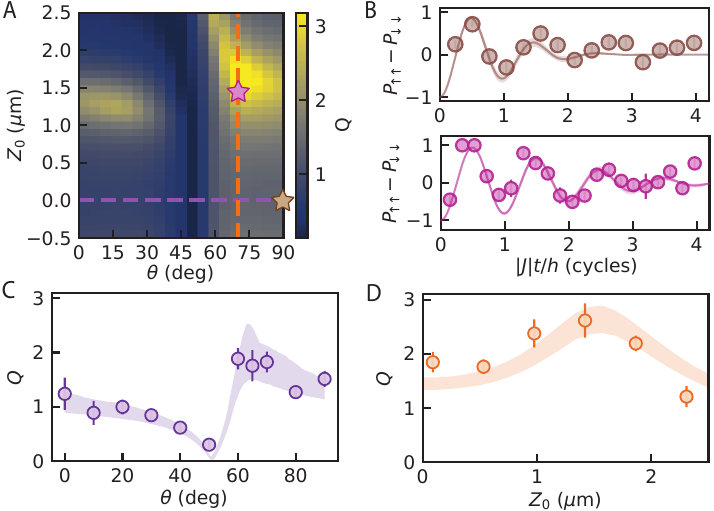}
   \caption{\textbf{Static geometric engineering of dipolar interactions}. (\textbf{A}) Predicted interaction quality factor $Q=|J|\tau/h$ over geometry space parameterized by $(\theta, z_0)$. The stars mark the geometries used for the dipolar oscillation measurements shown in B, while the horizontal (vertical) dashed lines mark the cuts depicted in C (D). (\textbf{B}) Comparison of dipolar oscillations at a reference point $(\theta, z_0) = (90^\circ, 0)$ and an optimal configuration $(\theta^*, z_0^*) = (70^\circ,1.42\,\mu$m). Solid lines are fit to a damped dipolar-exchange oscillation; corresponding fitted decay times are $\tau=$ 40(6)\,ms and 103(12)\,ms, respectively. (\textbf{C}) Dipolar interaction quality factor $Q$ as a function of angle $\theta$ at $z_0 = 0$. (\textbf{D}) Dipolar quality factor $Q$ as a function of $z_0$, at a fixed angle $\theta^* = 70^\circ$. Shaded bands in (C, D) show the two-body dipolar model simulations, with widths indicating uncertainty from finite numerical sampling.}
    \label{fig:fig2}
\end{figure}

\begin{figure*}[t!] \centering
\includegraphics[width=\textwidth]{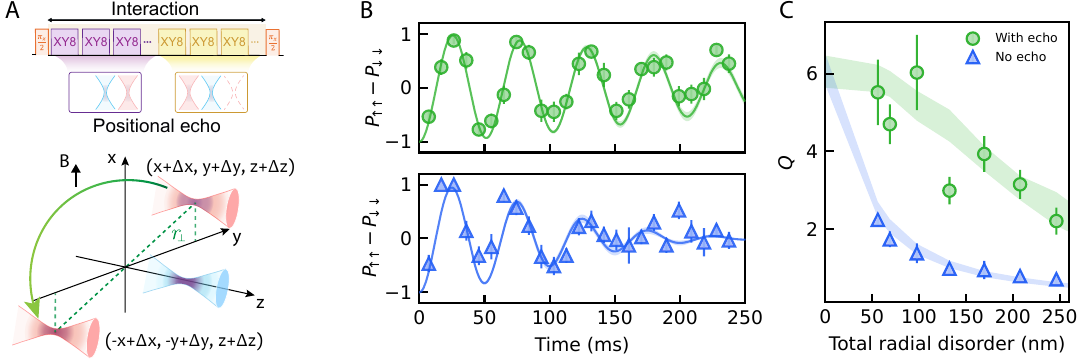}
   \caption{\textbf{Positional Echo Protocol.} (\textbf{A}) The interacting pair evolves for equal durations at two geometries, realized by rapidly moving one molecule midway through the interaction duration. The XY8 dynamical decoupling sequence is applied throughout the interaction. (\textbf{B}) Comparison of optimal dipolar oscillations with the motional echo (top, green) and without the motional echo (bottom, blue). (\textbf{C}) We characterize the performance of the echo protocol by injecting additional disorder into the SLM trap position. The total radial disorder is obtained by adding the injected disorder in quadrature with the uncorrected baseline radial disorder ($\simeq 50~\mathrm{nm}$); the first data points correspond to zero injected disorder. The echo protocol improves the dipolar coherence by a factor of 3.}
    \label{fig:fig3}
\end{figure*}

Guided by this geometric dependence, we next characterize the corresponding dipolar coherence across geometry space parametrized by $(\theta, z_0)$. We quantify this coherence by quality factor $Q \equiv |J|\tau /h$, where $\tau$ is the effective $1/e$ decay time of the spin-exchange oscillation contrast, including finite single-body coherence. Fig.~\ref{fig:fig2}A shows the predicted $Q$ for pairs of CaF molecules ($d \sim 1$\,Debye) with an in-plane separation of $r_\perp = 2.5~\mu$m. The two-body dipolar model is described in the supplementary material.

We first measure coherence in the planar configuration, with $z_0 = 0$. We vary the position of the AOD tweezers with respect to the static SLM tweezers, while maintaining a fixed magnetic field direction. The measured $Q$ for a pair of molecules exhibits a slight optimum near $\theta^{*} \approx 63^{\circ}$ (Fig.~\ref{fig:fig2}C), consistent with the previously proposed 2D magic angle~\cite{you2025control}, despite the reduced interaction strength. The lack of a sharply peaked feature reflects the trade-off between enhanced coherence and reduced interaction strength $J$. The corresponding two-qubit population dynamics $P_{\uparrow\uparrow}$ and $P_{\downarrow\downarrow}$ measured at $(z_0=0, \theta^{*})$, from which we extract the associated $\tau$, is shown in Fig.~\ref{fig:fig2}B. The experimental measurements reported in this work are taken from a single representative molecule pair to avoid axial disorder in the SLM array~\cite{chew2024ultraprecise} (Supplementary Text).

Having identified the optimal configuration within the planar geometry, we then vary the relative axial displacement $z_0$ at fixed $\theta^{*}$ by applying a variable offset of the defocus Zernike polynomial $Z^0_2$ on top of the base SLM phase hologram. Fig.~\ref{fig:fig2}D shows a peak in $Q$ at finite offset $z_0^{*} \approx 1.42\,\mu$m. At this 3D ``magic'' configuration, we measure a maximum $Q = 2.61(31)$. The 3D magic configuration improves by approximately a factor of 1.4 relative to the 2D magic-angle case, demonstrating that the additional geometric degree of freedom can substantially enhance interaction coherence by reducing the motional sensitivity of $J$ along all three axes. Representative dipolar spin-exchange dynamics measured at $(\theta^{*}, z_0^{*})$ are shown in Fig.~\ref{fig:fig2}B. 

\subsection*{Position noise and dynamic geometric echo}

While the interleaved SLM-AOD array allows for precise and flexible control of the pairwise geometry, any differential fluctuation in their paths perturbs the mean trap positions of an interacting pair, resulting in a first-order change to the intermolecular spacing. These positional fluctuations are slow compared to the timescale of the dipolar interaction but can differ between experimental realizations, thereby introducing a dephasing channel in the coherent dipolar interactions. This positional disorder becomes increasingly consequential as we approach a magic configuration, since the sensitivity to thermal fluctuations is reduced.
At these points, noise in relative position on the order of tens of nanometers can substantially limit the interaction coherence.

By imaging the two traps on a separate camera via a pick-off and feeding back on the tweezer position, we reduce the residual radial offset noise to approximately 50\,nm rms over 24 hours (see Supplementary Text for measurement details). At magic configurations described above, with reduced thermal sensitivity, these fluctuations become the dominant source of dipolar decoherence. The ability to dynamically tune the geometry of the interacting molecules, on timescales faster than their entanglement rate, therefore opens a route to mitigating the effects of positional disorder. 

We implement a two-step geometric echo that suppresses sensitivity to any static radial offsets. Since the relative position noise during the interaction is effectively static, it can be represented as a set of fixed offsets $(\Delta x, \Delta y, \Delta z)$. As illustrated in Fig.~\ref{fig:fig3}A, in the echo sequence, one molecule evolves for half of the sequence at position $(x+\Delta x,y+\Delta y, z +\Delta z)$ relative to the other molecule, and is then rapidly moved by the AODs to a symmetric geometry around $\hat{z}$ at $(-x+\Delta x,-y+\Delta y, z +\Delta z)$ for the second half of the interaction. In this way, the same positional disorder shifts the interaction strength in the opposite direction during the two equal segments, so any disorder-induced phase shifts cancel to first order while the desired interaction is preserved. In Fig.~\ref{fig:fig3}B, we compare dipolar oscillations taken at the separately optimized geometries with and without the echo protocol. Applying the echo protocol during dipolar interaction at the optimized 3D geometry $(\theta^{*} = 70^{\circ}, z_0^{*} = 1.8 ~\mu\text{m})$ yields a dipolar coherence with $Q = 6.03(97) $ (Fig.~\ref{fig:fig3}B).

We quantify the robustness of this geometric echo protocol by injecting random radial offsets to the SLM trap positions and comparing the interaction $Q$ factor with and without the geometric echo. We apply a different random grating on top of the SLM pattern every 10 seconds (see supplementary text). Fig.~\ref{fig:fig3}C shows the measured $Q$ as a function of the magnitude of the injected disorder. 
At any scale of disorder the echo performs dramatically better than without the echo. At sufficiently large displacements, the echo protocol does not cancel out higher-order effects, such as angle changes, and we see a degradation of the enhancement. This could be mitigated by using more complicated positional sequences~\cite{you2025control}. 
In our present implementation, the geometric echo only targets transverse positional disorder. We estimate the uncanceled relative motion along the axial direction to be at the $\sim 50$\,nm level and attribute it mainly to thermally induced drift in the differential beam path.

\subsection*{Bell state characterization}

Thus far, we have focused on using static and dynamic control over interaction geometry to suppress position fluctuations and reduce sensitivity to thermal motion. These techniques mitigate the first- and second-order sensitivities of the dipolar interaction to positional fluctuations. However, further cooling can still improve coherence by reducing any residual quadratic contribution, as well as higher-order perturbations~\cite{you2025control} that are not fully suppressed by geometric control. We combine these protocols with a partial Raman sideband cooling sequence~\cite{bao2024raman} and benchmark the fidelity of a generated Bell state.

\begin{figure*}[htbp!]\centering
\includegraphics{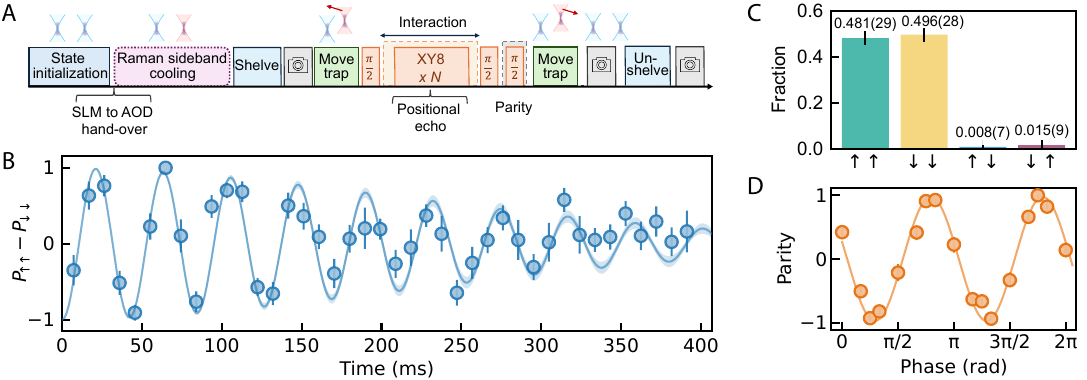}
    \caption{\textbf{Optimal Configuration and Bell state characterization.} (\textbf{A}) Illustration of the experimental sequence. Molecules in the SLM and AOD traps are simultaneously Raman-sideband cooled and then moved to the interaction geometry. After the interaction, a $\pi/2$ analysis pulse with variable phase $\phi$ is used for parity measurement, followed by detection of both the $\downstate$ and $\upstate$ states. (\textbf{B}) A representative oscillation with $Q=7.23(81)$. (\textbf{C}) Pair population statistics, measured in the computational basis at $t = 11.2$ ms of dipolar exchange evolution, corresponding to the $\pi/2$ point of the two-particle population oscillation. (\textbf{D}) Parity oscillations for a two-particle Bell state measured at the same evolution time. The fitted parity amplitude is $A_\Pi = 0.974^{+0.013}_{-0.018}$. Combining this with the pair populations in (C), we infer a Bell state fidelity of $\mathcal{F} = 0.976^{+0.008}_{-0.011}$.}
    \label{fig:fig4}
\end{figure*}

The experimental sequence is summarized schematically in Fig.~\ref{fig:fig4}A. After initializing all the molecules in $\upstate$, we apply partial Raman sideband cooling with the molecules already arranged and transferred to the interleaved array, as SLM-to-AOD trap handover after cooling can introduce substantial heating, especially at large axial displacements. To narrow the motional distribution while minimizing loss into weakly coupled motional states, we apply only a partial sideband cooling sequence, after which we estimate an axial occupation number of $\overline{n}_{z} \sim 10$ (Supplementary Text). As the final step of the cooling sequence also prepares the molecules in the $\downstate$ state, molecules are then directly moved to the target interaction geometry, where they evolve under dipolar exchange with the geometric echo applied. In Fig.~\ref{fig:fig4}B, we show the resulting dipolar spin-exchange dynamics of partially cooled molecules at $(\theta =70^\circ, z_0=1.42 ~\mu \text{m}, r_\perp= 1.9~\mu \text{m})$, yielding an interaction quality factor $Q=7.23(81)$, a $17\%$ improvement relative to without sideband cooling. Since the geometry is fundamentally insensitive to temperature, further cooling to the motional ground state is only expected to provide a modest gain in dipolar coherence. Remaining decoherence is dominated by uncorrected axial position noise, while additional cooling cycles incur additional population loss (see Supplementary Text). 

Using these experimental parameters, we generate a maximally entangled two-molecule Bell state by evolving the system for $1/4$ of the oscillation period~\cite{bao2023dipolar}. The state is characterized by the measured two-qubit populations and by a parity measurement, shown in Fig.~\ref{fig:fig4}C and D. After applying a $\pi/2$ pulse with variable phase $\phi$, we measure the parity $\Pi(\phi)= P_{\uparrow\uparrow} + P_{\downarrow\downarrow} - P_{\uparrow\downarrow} - P_{\downarrow\uparrow} $. A sinusoidal fit yields the parity oscillation amplitude $A_\Pi = 0.974^{+0.013}_{-0.018}$ (Fig.~\ref{fig:fig4}D). Combining this with the measured populations $P_{\uparrow\uparrow} + P_{\downarrow\downarrow}$ (Fig.~\ref{fig:fig4}C), we infer a fidelity of $\mathcal{F} = 0.976^{+0.008}_{-0.011}$, conditioned on molecules surviving through state preparation and partial sideband cooling. This fidelity is comparable to previous exemplary demonstrations of entanglement with assembled polar molecules near their motional ground state $(\overline{n}_z < 1)$~\cite{picard2025entanglement, ruttley2025long}, despite the molecules here having an axial occupation of $\overline{n}_{z} \sim 10$. The primary further limitations to improving dipolar coherence are threefold. First, the single-body coherence time is limited, partially due to black-body excitation between different vibrational states. Second, our current echo protocol does not cancel axial position disorder, which we estimate to be on the order of $\sim 50$\,nm over the timescale of this measurement. Dynamic tweezers in three dimensions~\cite{lu2025astigmatism, picard2025three}, a cryogenic environment~\cite{schymik2021single, zhang2025high}, and magic-wavelength trapping~\cite{gregory2024second} provide clear avenues for improvement. Third, improved cooling to $\overline{n}_z <1$ would further reduce higher-order thermal effects. Each of these improvements compounds, and all three of these in conjunction would result in Bell state fidelities of $\mathcal{F}\geq 0.999.$ Additional technical improvements and extensions of geometric control are discussed in the Supplementary Text. 

\subsection*{Outlook}
We have demonstrated that high-fidelity two-qubit entanglement of molecules can be reached through geometric control of the dipole-dipole interaction itself. The methods demonstrated here can be readily extended to other molecular tweezer platforms. Static three-dimensional geometry engineering reduces the sensitivity of the interaction to thermally induced motional broadening, while dynamic positional control refocuses slow relative positional noise between traps. This establishes a general strategy for suppressing quasi-static positional interaction disorder. Mitigating such disorder becomes increasingly important in larger arrays, where per-site calibration can be laborious, and when cooling lowers thermal broadening below the level of residual trap-position noise.

We have shown that a molecule can be moved during an entangling interaction without erasing the coherent dipolar evolution. Controlled motion thereby becomes not just a tool for assembly and rearrangement, but a corrective resource during the interaction. This places molecular tweezer arrays within a broader family of quantum processors in which motion is a coherent control resource. In trapped ions and neutral atoms, for example, transport has enabled reconfigurable connectivity while simply preserving quantum information~\cite{bluvstein2022quantum, pino2021demonstration, moses2023race}. Looking to the future, when geometry can be programmed on timescales that are short compared with the entangling dynamics, spatial control can be used as part of Hamiltonian engineering~\cite{you2025control}. This opens routes to robust two-qubit gates~\cite{webb2018resilient, bergonzoni2025iswap}, geometry-echo-protected many-body dynamics~\cite{choi2020robust}, and programmable molecular systems in which the network of effective couplings~\cite{bilitewski2023manipulating, katz2025floquet} can be dynamically engineered during coherent evolution.

\section*{Acknowledgments}
We are grateful to Wolfgang Ketterle for valuable discussions and helpful comments on the manuscript. This material is based upon work supported by the U.S. Department of Energy, Office of Science, National Quantum Information Science Research Centers, Quantum Systems Accelerator. Additional support is acknowledged from Harvard-MIT Center for Ultracold Atoms (Grant No. PHY-2317134); the Air Force Office of Scientific Research (AFOSR) AOARD under award number FA2386-24-1-4070; and from MURI W911NF-19-1-0283. S.S.Y. acknowledges support from the NSF GRFP. A.P. and S.S.Y. acknowledge support from the Harvard Quantum Initiative. Y.C. acknowledges support from the Quantum Information Research Support Center, funded by South Korea's Ministry of Science and ICT (MSIT) through the NRF of Korea (No. 2021M3H3A1036573). E.C. acknowledges support from the NRF of Korea (No. RS-2022-NR119745, RS-2024-00431938, RS-2024-00439981, and RS-2023-NR068116).

\bibliography{bib}

@article{burchesky2021rotational,
  title={Rotational coherence times of polar molecules in optical tweezers},
  author={Burchesky, Sean and Anderegg, Lo{\"\i}c and Bao, Yicheng and Yu, Scarlett S and Chae, Eunmi and Ketterle, Wolfgang and Ni, Kang-Kuen and Doyle, John M},
  journal={Physical Review Letters},
  volume={127},
  number={12},
  pages={123202},
  year={2021},
  publisher={APS}
}

@article{park2023extended,
  title={Extended rotational coherence of polar molecules in an elliptically polarized trap},
  author={Park, Annie J and Picard, Lewis RB and Patenotte, Gabriel E and Zhang, Jessie T and Rosenband, Till and Ni, Kang-Kuen},
  journal={Physical Review Letters},
  volume={131},
  number={18},
  pages={183401},
  year={2023},
  publisher={APS}
}

@article{gregory2024second,
  title={Second-scale rotational coherence and dipolar interactions in a gas of ultracold polar molecules},
  author={Gregory, Philip D and Fernley, Luke M and Tao, Albert Li and Bromley, Sarah L and Stepp, Jonathan and Zhang, Zewen and Kotochigova, Svetlana and Hazzard, Kaden RA and Cornish, Simon L},
  journal={Nature Physics},
  volume={20},
  number={3},
  pages={415--421},
  year={2024},
  publisher={Nature Publishing Group UK London}
}

@article{hepworth2025long,
  title={Long-lived multilevel coherences and spin-1 dynamics encoded in the rotational states of ultracold molecules},
  author={Hepworth, Tom R and Ruttley, Daniel K and von Gierke, Fritz and Gregory, Philip D and Guttridge, Alexander and Cornish, Simon L},
  journal={Nature Communications},
  volume={16},
  number={1},
  pages={7131},
  year={2025},
  publisher={Nature Publishing Group UK London}
}

@article{micheli2006toolbox,
  title={A toolbox for lattice-spin models with polar molecules},
  author={Micheli, Andrea and Brennen, Gavin K and Zoller, Peter},
  journal={Nature Physics},
  volume={2},
  number={5},
  pages={341--347},
  year={2006},
  publisher={Nature Publishing Group UK London}
}

@article{gadway2016strongly,
  title={Strongly interacting ultracold polar molecules},
  author={Gadway, Bryce and Yan, Bo},
  journal={Journal of Physics B: Atomic, Molecular and Optical Physics},
  volume={49},
  number={15},
  pages={152002},
  year={2016},
  publisher={IOP Publishing}
}

@article{gorshkov2011tunable,
  title={Tunable superfluidity and quantum magnetism with ultracold polar molecules},
  author={Gorshkov, Alexey V and Manmana, Salvatore R and Chen, Gang and Ye, Jun and Demler, Eugene and Lukin, Mikhail D and Rey, Ana Maria},
  journal={Physical Review Letters},
  volume={107},
  number={11},
  pages={115301},
  year={2011},
  publisher={APS}
}

@article{bilitewski2021dynamical,
  title={Dynamical generation of spin squeezing in ultracold dipolar molecules},
  author={Bilitewski, Thomas and De Marco, Luigi and Li, Jun-Ru and Matsuda, Kyle and Tobias, William G and Valtolina, Giacomo and Ye, Jun and Rey, Ana Maria},
  journal={Physical Review Letters},
  volume={126},
  number={11},
  pages={113401},
  year={2021},
  publisher={APS}
}

@article{bilitewski2023manipulating,
  title={Manipulating growth and propagation of correlations in dipolar multilayers: From pair production to bosonic Kitaev models},
  author={Bilitewski, Thomas and Rey, Ana Maria},
  journal={Physical Review Letters},
  volume={131},
  number={5},
  pages={053001},
  year={2023},
  publisher={APS}
}

@article{demille2002quantum,
  title={Quantum computation with trapped polar molecules},
  author={DeMille, David},
  journal={Physical Review Letters},
  volume={88},
  number={6},
  pages={067901},
  year={2002},
  publisher={APS}
}

@article{yelin2006dipolarQC,
  title={Schemes for robust quantum computation with polar molecules},
  author={Yelin, SF and Kirby, K and C{\^o}t{\'e}, Robin},
  journal={Physical Review A},
  volume={74},
  number={5},
  pages={050301},
  year={2006},
  publisher={APS}
}

@article{karra2016paramagnetic,
  title={Prospects for quantum computing with an array of ultracold polar paramagnetic molecules},
  author={Karra, Mallikarjun and Sharma, Ketan and Friedrich, Bretislav and Kais, Sabre and Herschbach, Dudley},
  journal={The Journal of chemical physics},
  volume={144},
  number={9},
  pages={094301},
  year={2016},
  publisher={AIP Publishing LLC}
}

@article{ni2018dipolar,
  title={Dipolar exchange quantum logic gate with polar molecules},
  author={Ni, Kang-Kuen and Rosenband, Till and Grimes, David D},
  journal={Chemical science},
  volume={9},
  number={33},
  pages={6830--6838},
  year={2018},
  publisher={Royal Society of Chemistry}
}

@article{sawant2020qudits,
  title={Ultracold polar molecules as qudits},
  author={Sawant, Rahul and Blackmore, Jacob A and Gregory, Philip D and Mur-Petit, Jordi and Jaksch, Dieter and Aldegunde, Jes{\'u}s and Hutson, Jeremy M and Tarbutt, MR and Cornish, Simon L},
  journal={New Journal of Physics},
  volume={22},
  number={1},
  pages={013027},
  year={2020},
  publisher={IOP Publishing}
}

@article{sundar2018synthetic,
  title={Synthetic dimensions in ultracold polar molecules},
  author={Sundar, Bhuvanesh and Gadway, Bryce and Hazzard, Kaden RA},
  journal={Scientific reports},
  volume={8},
  number={1},
  pages={3422},
  year={2018},
  publisher={Nature Publishing Group}
}

@article{cornish2024quantum,
  title={Quantum computation and quantum simulation with ultracold molecules},
  author={Cornish, Simon L and Tarbutt, Michael R and Hazzard, Kaden RA},
  journal={Nature Physics},
    volume={20},
  pages={730-740},
  year={2024},
  publisher={Nature Publishing Group UK London}
}

@article{kozyryev2017precision,
  title = {Precision Measurement of Time-Reversal Symmetry Violation with Laser-Cooled Polyatomic Molecules},
  author = {Kozyryev, Ivan and Hutzler, Nicholas R.},
  journal = {Phys. Rev. Lett.},
  volume = {119},
  issue = {13},
  pages = {133002},
  numpages = {6},
  year = {2017},
  month = {Sep},
  publisher = {American Physical Society},
  doi = {10.1103/PhysRevLett.119.133002},
  url = {https://link.aps.org/doi/10.1103/PhysRevLett.119.133002}
}

@article{kozyryev2021enhanced,
  title={Enhanced sensitivity to ultralight bosonic dark matter in the spectra of the linear radical SrOH},
  author={Kozyryev, Ivan and Lasner, Zack and Doyle, John M},
  journal={Physical Review A},
  volume={103},
  number={4},
  pages={043313},
  year={2021},
  publisher={APS}
}

@article{demille2024quantum,
  title={Quantum sensing and metrology for fundamental physics with molecules},
  author={DeMille, David and Hutzler, Nicholas R and Rey, Ana Maria and Zelevinsky, Tanya},
  journal={Nature Physics},
  volume={20},
  number={5},
  pages={741--749},
  year={2024},
  publisher={Nature Publishing Group UK London}
}

@article{hutzler2020polyatomic,
  title={Polyatomic molecules as quantum sensors for fundamental physics},
  author={Hutzler, Nicholas R},
  journal={Quantum Science and Technology},
  volume={5},
  number={4},
  pages={044011},
  year={2020},
  publisher={IOP Publishing}
}

@article{chupp2019electric,
  title={Electric dipole moments of atoms, molecules, nuclei, and particles},
  author={Chupp, TE and Fierlinger, Peter and Ramsey-Musolf, MJ and Singh, JT},
  journal={Reviews of Modern Physics},
  volume={91},
  number={1},
  pages={015001},
  year={2019},
  publisher={APS}
}

@article{cairncross2019atoms,
  title={Atoms and molecules in the search for time-reversal symmetry violation},
  author={Cairncross, William B and Ye, Jun},
  journal={Nature Reviews Physics},
  volume={1},
  number={8},
  pages={510--521},
  year={2019},
  publisher={Nature Publishing Group UK London}
}

@article{endres2016atom,
  title={Atom-by-atom assembly of defect-free one-dimensional cold atom arrays},
  author={Endres, Manuel and Bernien, Hannes and Keesling, Alexander and Levine, Harry and Anschuetz, Eric R and Krajenbrink, Alexandre and Senko, Crystal and Vuletic, Vladan and Greiner, Markus and Lukin, Mikhail D},
  journal={Science},
  volume={354},
  number={6315},
  pages={1024--1027},
  year={2016},
  publisher={American Association for the Advancement of Science}
}

@article{you2025control,
  title = {Control of Dipolar Dynamics by Geometrical Programming},
  author = {You, Jiaqi and Doyle, John M. and Liu, Zirui and Yu, Scarlett S. and Periwal, Avikar},
  journal = {Phys. Rev. Lett.},
  volume = {135},
  issue = {25},
  pages = {253002},
  numpages = {6},
  year = {2025},
  month = {Dec},
  publisher = {American Physical Society},
  doi = {10.1103/k22t-xgcy},
  url = {https://link.aps.org/doi/10.1103/k22t-xgcy}
}

@article{bao2023dipolar,
  title={Dipolar spin-exchange and entanglement between molecules in an optical tweezer array},
  author={Bao, Yicheng and Yu, Scarlett S and Anderegg, Lo{\"\i}c and Chae, Eunmi and Ketterle, Wolfgang and Ni, Kang-Kuen and Doyle, John M},
  journal={Science},
  volume={382},
  number={6675},
  pages={1138--1143},
  year={2023},
  publisher={American Association for the Advancement of Science}
}

@article{holland2023demand,
  title={On-demand entanglement of molecules in a reconfigurable optical tweezer array},
  author={Holland, Connor M and Lu, Yukai and Cheuk, Lawrence W},
  journal={Science},
  volume={382},
  number={6675},
  pages={1143--1147},
  year={2023},
  publisher={American Association for the Advancement of Science}
}

@article{picard2025entanglement,
  title={Entanglement and iSWAP gate between molecular qubits},
  author={Picard, Lewis RB and Park, Annie J and Patenotte, Gabriel E and Gebretsadkan, Samuel and Wellnitz, David and Rey, Ana Maria and Ni, Kang-Kuen},
  journal={Nature},
  volume={637},
  number={8047},
  pages={821--826},
  year={2025},
  publisher={Nature Publishing Group UK London}
}

@article{christakis2023probing,
  title={Probing site-resolved correlations in a spin system of ultracold molecules},
  author={Christakis, Lysander and Rosenberg, Jason S and Raj, Ravin and Chi, Sungjae and Morningstar, Alan and Huse, David A and Yan, Zoe Z and Bakr, Waseem S},
  journal={Nature},
  volume={614},
  number={7946},
  pages={64--69},
  year={2023},
  publisher={Nature Publishing Group UK London}
}

@article{ruttley2025long,
  title={Long-lived entanglement of molecules in magic-wavelength optical tweezers},
  author={Ruttley, Daniel K and Hepworth, Tom R and Guttridge, Alexander and Cornish, Simon L},
  journal={Nature},
  volume={637},
  number={8047},
  pages={827--832},
  year={2025},
  publisher={Nature Publishing Group UK London}
}

@article{bao2024raman,
  title={Raman sideband cooling of molecules in an optical tweezer array to the 3D motional ground state},
  author={Bao, Yicheng and Yu, Scarlett S and You, Jiaqi and Anderegg, Lo{\"\i}c and Chae, Eunmi and Ketterle, Wolfgang and Ni, Kang-Kuen and Doyle, John M},
  journal={Physical Review X},
  volume={14},
  number={3},
  pages={031002},
  year={2024},
  publisher={APS}
}

@article{picard2025three,
  title={A three-dimensional acousto-optic deflector},
  author={Picard, Lewis RB and Endres, Manuel},
  journal={arXiv preprint arXiv:2510.07633},
  year={2025}
}

@article{lu2025astigmatism,
  title={Astigmatism-free 3D optical tweezer control for rapid atom rearrangement},
  author={Lu, Yue-Hui and Song, Nathan and Xiang, Tai and Ho, Jacquelyn and Lee, Tsai-Chen and Yan, Zhenjie and Stamper-Kurn, Dan M},
  journal={arXiv preprint arXiv:2510.11451},
  year={2025}
}

@article{chew2024ultraprecise,
  title={Ultraprecise holographic optical tweezer array},
  author={Chew, Y Torii and Poitrinal, Martin and Tomita, Takafumi and Kitade, Sota and Mauricio, Jorge and Ohmori, Kenji and de L{\'e}s{\'e}leuc, Sylvain},
  journal={Physical Review A},
  volume={110},
  number={5},
  pages={053518},
  year={2024},
  publisher={APS}
}

@article{lu2026probing,
  title={Probing Coherent Many-Body Spin Dynamics in a Molecular Tweezer Array Quantum Simulator},
  author={Lu, Yukai and Holland, Connor M and Welsh, Callum L and Chen, Xing-Yan and Cheuk, Lawrence W},
  journal={arXiv preprint arXiv:2603.19090},
  year={2026}
}

@article{li2023tunable,
  title={Tunable itinerant spin dynamics with polar molecules},
  author={Li, Jun-Ru and Matsuda, Kyle and Miller, Calder and Carroll, Annette N and Tobias, William G and Higgins, Jacob S and Ye, Jun},
  journal={Nature},
  volume={614},
  number={7946},
  pages={70--74},
  year={2023},
  publisher={Nature Publishing Group UK London}
}

@article{yan2013observation,
  title={Observation of dipolar spin-exchange interactions with lattice-confined polar molecules},
  author={Yan, Bo and Moses, Steven A and Gadway, Bryce and Covey, Jacob P and Hazzard, Kaden RA and Rey, Ana Maria and Jin, Deborah S and Ye, Jun},
  journal={Nature},
  volume={501},
  number={7468},
  pages={521--525},
  year={2013},
  publisher={Nature Publishing Group UK London}
}

@article{hazzard2014many,
  title={Many-body dynamics of dipolar molecules in an optical lattice},
  author={Hazzard, Kaden RA and Gadway, Bryce and Foss-Feig, Michael and Yan, Bo and Moses, Steven A and Covey, Jacob P and Yao, Norman Y and Lukin, Mikhail D and Ye, Jun and Jin, Deborah S and others},
  journal={Physical Review Letters},
  volume={113},
  number={19},
  pages={195302},
  year={2014},
  publisher={APS}
}

@article{li2021tuning,
  title={Tuning of dipolar interactions and evaporative cooling in a three-dimensional molecular quantum gas},
  author={Li, Jun-Ru and Tobias, William G and Matsuda, Kyle and Miller, Calder and Valtolina, Giacomo and De Marco, Luigi and Wang, Reuben RW and Lassabli{\`e}re, Lucas and Qu{\'e}m{\'e}ner, Goulven and Bohn, John L and others},
  journal={Nature Physics},
  volume={17},
  number={10},
  pages={1144--1148},
  year={2021},
  publisher={Nature Publishing Group UK London}
}

@article{miller2024two,
  title={Two-axis twisting using Floquet-engineered XYZ spin models with polar molecules},
  author={Miller, Calder and Carroll, Annette N and Lin, Junyu and Hirzler, Henrik and Gao, Haoyang and Zhou, Hengyun and Lukin, Mikhail D and Ye, Jun},
  journal={Nature},
  volume={633},
  number={8029},
  pages={332--337},
  year={2024},
  publisher={Nature Publishing Group UK London}
}

@article{carroll2025observation,
  title={Observation of generalized tJ spin dynamics with tunable dipolar interactions},
  author={Carroll, Annette N and Hirzler, Henrik and Miller, Calder and Wellnitz, David and Muleady, Sean R and Lin, Junyu and Zamarski, Krzysztof P and Wang, Reuben RW and Bohn, John L and Rey, Ana Maria and others},
  journal={Science},
  volume={388},
  number={6745},
  pages={381--386},
  year={2025},
  publisher={American Association for the Advancement of Science}
}

@article{choi2020robust,
  title={Robust dynamic Hamiltonian engineering of many-body spin systems},
  author={Choi, Joonhee and Zhou, Hengyun and Knowles, Helena S and Landig, Renate and Choi, Soonwon and Lukin, Mikhail D},
  journal={Physical Review X},
  volume={10},
  number={3},
  pages={031002},
  year={2020},
  publisher={APS}
}

@article{katz2025floquet,
  title={Floquet control of interactions and edge states in a programmable quantum simulator},
  author={Katz, Or and Feng, Lei and Porras, Diego and Monroe, Christopher},
  journal={Nature Communications},
  volume={16},
  number={1},
  pages={8815},
  year={2025},
  publisher={Nature Publishing Group UK London}
}

@article{webb2018resilient,
  title={Resilient entangling gates for trapped ions},
  author={Webb, Anna E and Webster, Simon C and Collingbourne, S and Bretaud, David and Lawrence, Adam M and Weidt, Sebastian and Mintert, Florian and Hensinger, Winfried K},
  journal={Physical Review Letters},
  volume={121},
  number={18},
  pages={180501},
  year={2018},
  publisher={APS}
}

@article{yu2024conveyor,
  title={A conveyor-belt magneto-optical trap of CaF},
  author={Yu, Scarlett S and You, Jiaqi and Bao, Yicheng and Anderegg, Lo{\"\i}c and Hallas, Christian and Li, Grace K and Lim, Dongkyu and Chae, Eunmi and Ketterle, Wolfgang and Ni, Kang-Kuen and others},
  journal={arXiv preprint arXiv:2409.15262},
  year={2024}
}

@article{bao2022fast,
  title={Fast optical transport of ultracold molecules over long distances},
  author={Bao, Yicheng and Yu, Scarlett S and Anderegg, Lo{\"\i}c and Burchesky, Sean and Gonzalez-Acevedo, Derick and Chae, Eunmi and Ketterle, Wolfgang and Ni, Kang-Kuen and Doyle, John M},
  journal={New Journal of Physics},
  volume={24},
  number={9},
  pages={093028},
  year={2022},
  publisher={IOP Publishing}
}

@article{cheuk2018lambda,
  title={$\Lambda$-enhanced imaging of molecules in an optical trap},
  author={Cheuk, Lawrence W and Anderegg, Lo{\"\i}c and Augenbraun, Benjamin L and Bao, Yicheng and Burchesky, Sean and Ketterle, Wolfgang and Doyle, John M},
  journal={Physical Review Letters},
  volume={121},
  number={8},
  pages={083201},
  year={2018},
  publisher={APS}
}

@article{schymik2021single,
  title={Single atoms with 6000-second trapping lifetimes in optical-tweezer arrays at cryogenic temperatures},
  author={Schymik, Kai-Niklas and Pancaldi, Sara and Nogrette, Florence and Barredo, Daniel and Paris, Julien and Browaeys, Antoine and Lahaye, Thierry},
  journal={Physical Review Applied},
  volume={16},
  number={3},
  pages={034013},
  year={2021},
  publisher={APS}
}

@article{zhang2025high,
  title={High optical access cryogenic system for Rydberg atom arrays with a 3000-second trap lifetime},
  author={Zhang, Zhenpu and Hsu, Ting-Wei and Tan, Ting You and Slichter, Daniel H and Kaufman, Adam M and Marinelli, Matteo and Regal, Cindy A},
  journal={PRX Quantum},
  volume={6},
  number={2},
  pages={020337},
  year={2025},
  publisher={APS}
}

@article{bluvstein2022quantum,
  title={A quantum processor based on coherent transport of entangled atom arrays},
  author={Bluvstein, Dolev and Levine, Harry and Semeghini, Giulia and Wang, Tout T and Ebadi, Sepehr and Kalinowski, Marcin and Keesling, Alexander and Maskara, Nishad and Pichler, Hannes and Greiner, Markus and others},
  journal={Nature},
  volume={604},
  number={7906},
  pages={451--456},
  year={2022},
  publisher={Nature Publishing Group UK London}
}

@article{moses2023race,
  title={A race-track trapped-ion quantum processor},
  author={Moses, Steven A and Baldwin, Charles H and Allman, Michael S and Ancona, R and Ascarrunz, L and Barnes, C and Bartolotta, J and Bjork, B and Blanchard, P and Bohn, M and others},
  journal={Physical Review X},
  volume={13},
  number={4},
  pages={041052},
  year={2023},
  publisher={APS}
}

@article{pino2021demonstration,
  title={Demonstration of the trapped-ion quantum CCD computer architecture},
  author={Pino, Juan M and Dreiling, Jennifer M and Figgatt, Caroline and Gaebler, John P and Moses, Steven A and Allman, MS and Baldwin, CH and Foss-Feig, Michael and Hayes, David and Mayer, Karl and others},
  journal={Nature},
  volume={592},
  number={7853},
  pages={209--213},
  year={2021},
  publisher={Nature Publishing Group UK London}
}

@article{bergonzoni2025iswap,
  title={iSWAP gate with polar molecules: Robustness criteria for entangling operations},
  author={Bergonzoni, Matteo and Jandura, Sven and Pupillo, Guido},
  journal={Physical Review A},
  volume={112},
  number={3},
  pages={032621},
  year={2025},
  publisher={APS}
}

@article{machu2025full,
  title={Full-field-of-view aberration correction for large arrays of focused beams},
  author={Machu, Yohann and Creutzer, Gautier and Sayrin, Cl{\'e}ment and Brune, Michel},
  journal={arXiv preprint arXiv:2512.15967},
  year={2025}
}

@article{christen2025full,
  title={Full-volume aberration-space holography},
  author={Christen, Ian and Panuski, Christopher and Propson, Thomas and Englund, Dirk},
  journal={arXiv preprint arXiv:2505.08777},
  year={2025}
}

@article{kim2019large,
  title={Large-scale uniform optical focus array generation with a phase spatial light modulator},
  author={Kim, Donggyu and Keesling, Alexander and Omran, Ahmed and Levine, Harry and Bernien, Hannes and Greiner, Markus and Lukin, Mikhail D and Englund, Dirk R},
  journal={Optics letters},
  volume={44},
  number={12},
  pages={3178--3181},
  year={2019},
  publisher={Optical Society of America}
}

@article{holland2024demonstration,
  title={Demonstration of erasure conversion in a molecular tweezer array},
  author={Holland, Connor M and Lu, Yukai and Li, Samuel J and Welsh, Callum L and Cheuk, Lawrence W},
  journal={arXiv preprint arXiv:2406.02391},
  year={2024}
}

@article{wu2023millisecond,
  title={Millisecond-lived circular Rydberg atoms in a room-temperature experiment},
  author={Wu, Haiteng and Richaud, R{\'e}mi and Raimond, J-M and Brune, Michel and Gleyzes, S{\'e}bastien},
  journal={Physical Review Letters},
  volume={130},
  number={2},
  pages={023202},
  year={2023},
  publisher={APS}
}

@article{park2017second,
  title={Second-scale nuclear spin coherence time of ultracold 23Na40K molecules},
  author={Park, Jee Woo and Yan, Zoe Z and Loh, Huanqian and Will, Sebastian A and Zwierlein, Martin W},
  journal={Science},
  volume={357},
  number={6349},
  pages={372--375},
  year={2017},
  publisher={American Association for the Advancement of Science}
}

@article{emperauger2025benchmarking,
  title={Benchmarking direct and indirect dipolar spin-exchange interactions between two Rydberg atoms},
  author={Emperauger, Gabriel and Qiao, Mu and Bornet, Guillaume and Chen, Cheng and Martin, Romain and Chew, Yuki Torii and G{\'e}ly, Bastien and Klein, Lukas and Barredo, Daniel and Browaeys, Antoine and others},
  journal={Physical Review A},
  volume={111},
  number={6},
  pages={062806},
  year={2025},
  publisher={APS}
}
\clearpage

\onecolumngrid
\begin{center}
\textbf{\large Supplementary Information}
\end{center}
\setcounter{section}{0}
\setcounter{figure}{0}
\setcounter{table}{0}
\renewcommand{\thesection}{S\arabic{section}}
\renewcommand{\thefigure}{S\arabic{figure}}
\renewcommand{\thetable}{S\arabic{table}}

\ifSubfilesClassLoaded{%
  \title{Supplementary Material}
  \maketitle
}{}
\setcounter{figure}{0}
\setcounter{secnumdepth}{1}
\renewcommand{\thefigure}{S\arabic{figure}}
\onecolumngrid
\section{Experimental Parameters}
\subsection{Molecule Loading and State Preparation}
CaF (calcium monofluoride) molecules are first captured in a red-detuned rf magneto-optical trap (MOT) and then transferred to a higher-density, blue-detuned conveyor-belt MOT. This facilitates loading of CaF molecules into a moving lattice, which transports the molecules into a separate glass cell, where they are loaded using $\Lambda$-enhanced gray molasses ($\Lambda$-imaging light) into a $4\times 8$ array of static tweezers generated by a spatial light modulator. Details for these procedures can be found in previous experiments~\cite{cheuk2018lambda, bao2022fast, bao2023dipolar, yu2024conveyor}.

Once the molecules are loaded into the SLM tweezers, we take a nondestructive image using $\Lambda$-imaging~\cite{cheuk2018lambda}. We then use dynamic tweezers generated by crossed acousto-optic deflectors (AODs) to rearrange molecules into a set of spatially alternating pairs. We alternate pairs in order to minimize any many-body effects, such that any pair-to-pair interaction rate is geometrically suppressed.

After rearrangement, we reduce the SLM tweezer trap depth to $k_B \times$ 400\,$\mu$K and optically pump the molecules into the $\lvert{v = 0, N = 1, F = 2, m_F = -2}\rangle$ state. We use a pair of composite microwave pulses to transfer population into the $\upstate = \lvert{v = 0, N = 1, F = 0, m_F = 0}\rangle$ state through the $\lvert{v = 0, N = 0, F = 1, m_F = -1}\rangle$ state. Between the two microwave pulses we resonantly excite any remaining molecules in $\lvert{N = 1}\rangle$ to heat them out of the tweezer. This sequence prepares pairs of molecules in $\upstate$ in the SLM array.

We then ramp on a $4\times 4$ array of dynamic AOD tweezers, which are overlapped with every other site of the SLM array over 250 $\mu$s (see Fig.~1). The AOD tweezers are 3--4 times deeper than the SLM traps to ensure high transfer fidelity, even with a large axial displacement between the two generated potentials. The transfer fidelity between the AOD and SLM traps is $> 99.9\%$ when the two tweezers share the same focal plane, and decreases to $95\%$ when their relative shift in the focal plane is larger than 2.5\,$\mu$m. However, we find that this infidelity primarily results in heating and loss as the AOD tweezer leaves the SLM tweezer, rather than a molecule remaining in the SLM tweezer.

For data taken with Raman sideband cooling, the loaded AOD traps are moved away from the SLM traps in 800\,$\mu$s toward an intermediate position for Raman sideband cooling. After cooling, the AOD tweezers are moved to the target interaction position. 

\subsection{Interaction and Detection}
Once the two molecules are prepared in $\lvert{\uparrow\uparrow}\rangle$ and in the desired geometry, we probe the interaction dynamics with microwave Ramsey spectroscopy on the $\upstate-\downstate$ transition. During the interaction time, we apply an XY8 dynamical decoupling sequence, with an interpulse time of 200\,$\mu$s, and a $\pi$ pulse duration of $8.5$\,$\mu$s. Each pulse has a Blackman profile, and the microwave horn is aligned to maximize the $\pi$-polarized component in order to suppress coupling to unwanted states. The XY8 decoupling sequence mitigates dephasing from magnetic field fluctuations, measured to be $70$\,$\mu$G, and from differential light shifts induced by the tweezer light~\cite{burchesky2021rotational}. With this decoupling sequence, we measure a single-body coherence time $T_2$ of $\tau_{sb} = 0.981(23)$\,s (Fig.~\ref{fig:sb_coherence}). This time is comparable for both the AOD and SLM tweezers. Furthermore, we find that the optimal XY8 parameters and the measured single-particle coherence do not depend appreciably on the interaction geometry.  

For our echo protocol, we move the molecules in AOD tweezers along a three-segment trajectory. First, with an AOD tweezer at interaction position $(x, y, z)$, we move the tweezer away from the SLM-trapped molecule to minimize any effects of the dipolar interaction, moving to $(x + 2y, y - 2x, z)$. We then move to $(-x + 2y, -y-2x, z)$, and finally finish the echo by moving to $(-x, -y, z)$. We make each one of these moves with a constant-jerk trajectory over the course of $800$\,$\mu$s, so that the total echo takes 2.4\,ms. This timing aligns the move around the $\pi$ pulses in the dynamical decoupling sequence, and we observe no loss in single-body coherence with and without the echo protocol. Furthermore, these moves happen on length scales of a few microns, and we find no molecule loss even at reduced echo durations of 100\,$\mu$s.

For measurements of the interaction quality factor $Q$, we extract the Ramsey contrast after the interaction by applying a single $\pi/2$ pulse to read out in the computational basis. For parity measurements of the generated Bell state, we apply an additional $\pi/2$ pulse with variable phase and measure the resulting parity oscillation.
\begin{figure}[!h]
    \centering
    \includegraphics[width=0.45\columnwidth]{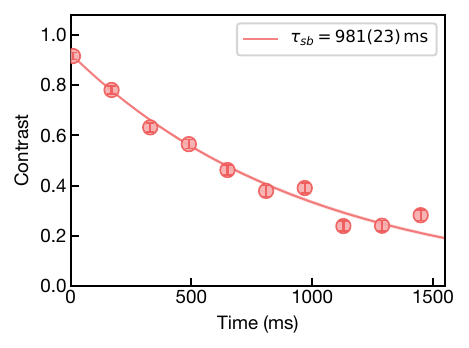}
    \caption{Single-body coherence of trapped molecules. Extracted Ramsey contrast is shown as a function of interrogation time with an XY8 dynamical decoupling sequence. Solid curve shows an exponential decay fit $e^{-t/\tau_{sb}}$, giving single-body coherence $\tau_{sb} = 0.981(23)$\,s. } 
    \label{fig:sb_coherence}
\end{figure}
After the final microwave pulse, we move the AOD tweezers back to their original positions, so that they overlap the SLM traps, and ramp them down, thus returning all molecules to the SLM-generated tweezers. The $\Lambda$-imaging light addresses molecules in $\upstate$. We first image molecules in the $N=1$ rotational state, then apply a resonant heating pulse to remove all molecules remaining in $N=1$. Finally, we apply a composite microwave $\pi$ pulse to transfer population from $\downstate$ to $\upstate$ and image again~\cite{bao2023dipolar}. All data are conditioned on retaining both molecules of an interacting pair. 

\section{Two-molecule dipolar interaction model}
At the temperatures relevant to the experiment, the spatial extent of the trapped molecular wavefunctions is not negligible compared with the length scale over which the dipole-dipole interaction varies. We therefore evaluate the dipolar exchange matrix element using the quantized motional wavefunctions of the two molecules, and then average over the thermal and static positional distributions sampled in the experiment.
For a pair of molecules with motional states $\mathbf n_i=(n_{ix},n_{iy},n_{iz})$, we write the dipolar spin-exchange matrix element as

\begin{equation}
\frac{J(\bf{n_1},\bf{n_2};\bf{r})}{2}
=
-\frac{d_{\downarrow\uparrow}^2}{4\pi\epsilon_0}
\left\langle \bf{n_1}\bf{n_2} \middle|
\frac{1-3(\hat{B}\cdot\bf{r})^2/r^2}{r^3}
\middle| \bf{n_1}\bf{n_2} \right\rangle ,
\end{equation}
where $d_{\downarrow\uparrow}$ is the transition dipole moment,
$\hat{B}$ is the quantization-axis direction, and
$\mathbf r=\bf{r_1} -\bf{r_2}+\bf{r_0}$ is the molecular separation
including the nominal tweezer displacement $\bf{r_0}$ (i.e., trap positions without shot-to-shot position disorder, as discussed below). Each $|\bf {n_i}\rangle$ is taken to be a product of one-dimensional harmonic oscillator eigenstates with the measured trap frequencies $(\omega_x,\omega_y,\omega_z)$.
We parameterize the nominal relative displacement $\bf{r_0}$ by the in-plane angle $\theta$ and axial offset $z_0$ as
\begin{equation}
\mathbf r_0=(r_\perp\cos\theta,\,r_\perp\sin\theta,\,z_0),
\end{equation}

To obtain the predicted dipolar coherence time and interaction strength for a given geometry, we perform a Monte Carlo average over motional occupation and static relative-position disorder. The motional quantum numbers for the two molecules are drawn independently along each axis $\alpha \in \{x,y,z\}$ from Boltzmann thermal distributions $p_\alpha(n) \propto e^{\frac{-n\hbar\omega_\alpha}{k_\mathrm{B}T}} $, where the temperature $T$ is specified by the mean occupation $\bar n$: $k_\mathrm{B}T=\hbar\omega_z/\ln(1+1/\bar n)$.

In addition to motional spread, we include shot-to-shot relative position noise between the AOD and SLM trap centers, arising from fluctuations in the differential beam paths. We model this noise as a static displacement drawn from
\begin{equation}
(\delta x,\delta y,\delta z) \sim
\mathcal N(0,\sigma_x^2)
\times
\mathcal N(0,\sigma_y^2)
\times
\mathcal N(0,\sigma_z^2),
\end{equation}
with $\sigma_x=\sigma_y=\sigma_\perp$. The sampled relative displacement is therefore $\mathbf r_s=(x_0+\delta x,y_0+\delta y,z_0+\delta z)$. 

For each Monte Carlo sample $s$, the exchange matrix element is numerically evaluated and converted to an exchange
frequency $f_s=J_s/h$. Repeating over $N_\text{samples}$ draws of $(\mathbf{n}_1, \mathbf{n}_2)$ and $(\delta x, \delta y, \delta z)$ yields the coupling distribution $P(f)$ for that geometry. The Fourier transform of $P(f)$ gives the ensemble-averaged two-body contrast $C(t)=\left|\langle e^{i2\pi f_s t}\rangle_s\right|.$ 

We extract the simulated coherence time $\tau$ as the time at which the envelope, including the same single-molecule decay $e^{-t/\tau_{sb}}$, first decays to $1/e$. The simulated quality factor is then
\begin{equation}
Q=|\bar{f}|\,\tau,
\qquad
\bar{f}=\frac{1}{N_\mathrm{samples}}\sum_s f_s .
\end{equation}

To implement the positional echo protocol, the configuration during the second half of the interaction reverses the in-plane displacement $\bf r_\perp \rightarrow - \bf r_\perp$. The effective coupling for each sample is the time-weighted average of the couplings in the different geometric configurations.

The shaded simulation bands shown in the main text quantify the simulation numerical sampling uncertainty. At each plotted geometry, we bootstrap-resample the coupling ensemble $\{f_s\}$ with replacement, recompute $Q$ for each resampled ensemble, and plot the central 68\% interval.

For the experimental data and simulation results, the quality factor $Q$ reported in the main text is calculated from the fitted oscillation frequency $f$ and effective $1/e$ decay time $\tau$ of the oscillation envelope. The fit model used for the experimental oscillation envelope is $D(t)=e^{-t/\tau_{sb}} e^{-(t/\tau_{tb})^2}$, where $\tau_{tb}$ is the two-body coherence. Thus, $\tau$ is defined by $D(\tau) =1/e$, or equivalently $(\tau/\tau_{sb})+(\tau/\tau_{tb})^2=1$.

\section{Raman sideband cooling}
We perform Raman sideband cooling using a scheme based on our past demonstration~\cite{bao2024raman}, with modifications optimized for interaction coherence rather than for maximizing the motional ground-state fraction alone.

In this study, the relevant figure of merit is dipolar interaction coherence. Thermal dephasing of the dipolar interaction is set by the width of the sampled motional distribution. In addition, slow relative focal shifts between the traps can broaden the effective interaction coupling if each data point requires long averaging. A higher number of surviving molecules after Raman sideband cooling reduces the acquisition time needed for a given statistical precision, thereby reducing sensitivity to slow differential focal drift. We therefore use a partial cooling sequence that narrows the resultant motional distribution while maintaining a higher molecular survival probability.
 
Since the CaF molecules trapped in tweezers are initially far outside the Lamb–Dicke regime, a fixed-order sideband-cooling sequence can leave population in motional states that are only weakly coupled to the selected Raman transition. These motional dark states can leave a high-energy tail in the final motional distribution even as the population near the ground state increases. We begin with higher-order axial sideband cooling pulses, $\Delta n_z=10$, to remove a large number of motional quanta along the weakly confined $z$ axis. As the axial distribution narrows, we progressively reduce $\Delta n_z$ while sparsely interspersing radial cooling pulses along $x$ and $y$. We terminate the sequence at $\Delta n_z = 5$ and $\Delta n_{x,y} = 2$. This partial cooling sequence balances the reduction of motional width relevant for dipolar dephasing against molecule survival.

We estimate the final motional occupation using Monte Carlo simulations of the cooling sequence. For the axial direction, the simulated distribution peaks at $n_z = 6$, with 72\% of the population occupying states with $n_z < 12$.

To more directly probe the effect of sideband cooling, we measure the dipolar interaction frequency as a function of relative axial displacement $z_0$ between two molecules at an angle of $\theta = 0$. We observe the population $P(z_0) \sim \cos(\phi(z_0))$ and convert the fitted phase to interaction frequency through $f(z_0) = \phi(z_0)/2\pi t_\text{int} $.  
The data in Fig.~\ref{fig:cooling} show two clear signatures of reduced motional excitations: partial Raman sideband cooling increases the interaction frequency and narrows the axial response profile. This measurement also provides intuition for the magic condition, which occurs roughly at the point where the two curves intersect. At this point the interaction rate has reduced sensitivity to the underlying motional distribution.
\begin{figure}[!h]
    \centering
    \includegraphics[width=0.45\columnwidth]{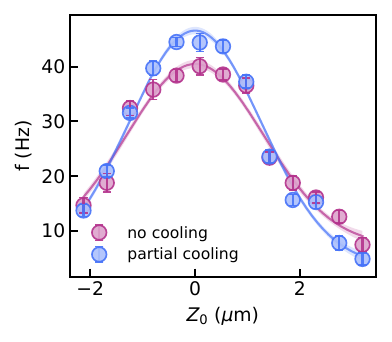}
    \caption{Interaction frequency as a function of axial displacement, with and without partial sideband cooling applied. } 
    \label{fig:cooling}
\end{figure}

\section{Homogenization of Tweezer Arrays}
The dipolar interaction is strongly sensitive to the relative positions of the two molecules. In our experiment,
these positions are set by the two independently propagated beam paths for the SLM and AOD tweezer arrays. Therefore, the relevant quantity is the relative trap-center alignment between the two arrays. Slow differential drift between them produces shot-to-shot disorder in the interaction strength, making precise alignment and calibration essential.

To monitor the relative transverse alignment, we collect the leaked light from a back-polished final mirror before the objective and image the tweezer light on a camera. This image provides the site-resolved positions of the SLM and AOD tweezers in the $x$-$y$ plane. We use these measured positions to calibrate the SLM hologram and align each SLM-generated tweezer to the corresponding AOD-generated tweezer. The correction is obtained through repeated iterations of a weighted Gerchberg--Saxton algorithm~\cite{kim2019large} and is then held fixed during data taking. We independently check the site-resolved alignment by optimizing the handoff fidelity between SLM and AOD tweezers, and find good agreement with the camera-based calibration within our measurement precision.

The dipolar interaction is also sensitive to relative axial displacement. Without compensation, SLM-generated tweezer arrays can exhibit site-dependent axial offsets exceeding 100\,nm~\cite{chew2024ultraprecise}. We calibrate these offsets by first measuring the AOD--SLM handoff fidelity as a function of the defocus Zernike mode $Z^0_2$ for each interacting pair. We then apply site-dependent defocus corrections to the SLM hologram, reducing the residual trap-to-trap axial disorder to less than 100\,nm rms~\cite{chew2024ultraprecise, christen2025full, machu2025full}. We further confirm this by using the dipolar interaction strength to measure the relative focal shift between a pair of SLM and AOD tweezers, as shown in Fig.~\ref{fig:cooling}.

Finally, we homogenize the trap depths for single-body coherence across the array, as the $\upstate \longleftrightarrow \downstate$ transition frequency depends on the trap intensity and polarization~\cite{burchesky2021rotational}. For the SLM array, we rearrange molecules into the SLM-generated tweezers and perform Ramsey spectroscopy on the $\upstate \longleftrightarrow \downstate$ transition, and apply site-dependent hologram corrections. For the AOD array, we perform the same compensation with molecules rearranged into the AOD tweezers, and compensate the rf intensities for each row and column. In principle, these three compensations (shifts in transverse plane, focal shifts, and intensity compensations) are independent for the SLM. However, changes in the hologram can weakly affect these different compensations. We therefore iterate through each of these measurements and compensations. Once calibrated, the arrays can remain stable for multiple months, provided that the thermal load on the AODs and SLM is stable.

Once the SLM hologram and AOD intensities are fixed, we are still sensitive to slow drifts of the two beam paths. During experiments, we correct residual slow, global drift between the SLM and AOD beam paths using the pickoff-camera signal. We image both tweezer arrays during the interaction sequence and extract the mean relative displacement between them. This average offset is compensated by applying a global grating to the SLM hologram. The grating is updated every two minutes, and whenever an experimental parameter, such as the interaction duration, is changed.

\section{Future avenues for improving entanglement fidelity}
The Bell-state fidelity reported in this work ($\mathcal{F} = 0.976^{+0.008}_{-0.011}$, with an oscillation quality factor $Q \approx 7.2$) is limited by a combination of single-body decoherence, residual thermal motion, and relative tweezer position disorder. Here we outline several upgrades, each addressing a distinct one of these limitations.

\textit{Axial position disorder.} After incorporating our echo protocol, shot-to-shot axial tweezer position fluctuations, on the order of $\sigma_z \sim 40$~nm, are one of the dominant sources of interaction decoherence. The present geometric echo in our experiment refocuses radial position noise, but does not cancel axial fluctuations. Implementing the echo with a three-dimensional AOD setup~\cite{lu2025astigmatism, picard2025three} would allow $\sigma_z$ to be refocused as well, suppressing this contribution. Improved passive stabilization of the tweezer optics is a simpler intermediate step that can be pursued in parallel. 
The optimal interaction geometry also depends strongly on the residual axial disorder. Simulations at $\bar{n}_z = 10$ show that the present $\theta = 70^\circ$, $r_{\perp} = 2~\mu$m, $z_0=1.8~\mu$m configuration is more robust when $\sigma_z$ is large. 
Once effective axial fluctuations are reduced, either through passive stability or a three-dimensional echo, it becomes possible to work at another magic-angle condition, near $\theta = 20^\circ$ (see Fig.~1C), which has a larger interaction strength. For the same single-particle coherence, we expect this condition to increase $Q$ by a factor of two compared to what we report here (see Fig.~\ref{fig:sm_cooling_sigma_z}).

\begin{figure}[htbp]
    \centering
    \includegraphics[width=0.5\columnwidth]{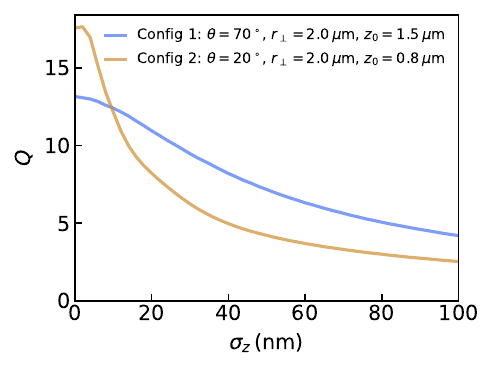}
    \caption{
    Simulated oscillation quality factor $Q$ versus axial position disorder $\sigma_z$ for two interaction geometries at $\bar{n}_z=10$. The $\theta=70^\circ$, $r_{\perp} =2~\mu$m, $z_0=1.8~\mu$m geometry is more robust to large axial disorder, while the $\theta=20^\circ$, $r_{\perp}=2~\mu$m, $z_0=0.8~\mu$m geometry gives a larger $Q$ when $\sigma_z \lesssim 20$~nm because of its stronger dipolar interaction.
    }
    \label{fig:sm_cooling_sigma_z}
\end{figure}

\textit{Reduced thermal occupation.} Once positional disorder is suppressed below the thermal extent of the molecular wave packet, residual thermal occupation of the motional modes becomes the leading source of interaction-strength disorder. This remains true even at the magic configuration, because the static magic-angle condition cancels the sensitivity to thermal motion only to leading order. 
Simulations show that additional cooling is most beneficial when axial disorder is already well controlled; otherwise, the residual axial disorder dominates and obscures the improvement from reducing $\bar{n}$ (Fig.~\ref{fig:sm_cooling_nbar}). 
Additional cooling toward the motional ground state~\cite{bao2024raman} is therefore expected to further increase $Q$.

\begin{figure}[htbp]
    \centering
    \includegraphics[width=0.5\columnwidth]{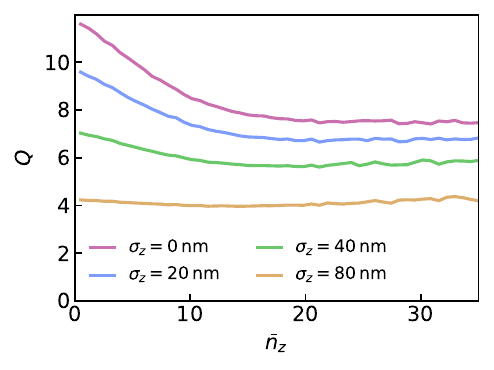}
    \caption{\textbf{Simulated effect of reducing thermal occupation.}
    Simulated interaction quality factor $Q$ as a function of mean motional occupation $\bar n$ for different rms axial position disorders $\sigma_z$. The simulation uses the two-step geometric echo configuration
    $(r_\perp,\theta,z_0)=(2\,\mu\mathrm{m},70^\circ,1.8\,\mu\mathrm{m})$.
    Reducing $\bar n_z$ through Raman sideband cooling increases $Q$ only when axial disorder is sufficiently well controlled.
    For larger $\sigma_z$, the spread in the mean relative axial position dominates the disorder in $J$, so further cooling gives a smaller improvement.
    }
    \label{fig:sm_cooling_nbar}
\end{figure}

\textit{Tighter interaction geometries.} Reducing the interparticle spacing increases the dipolar coupling, but it also amplifies sensitivity to positional disorder. Once both tweezer position disorder and thermal motion are sufficiently suppressed, tighter geometries provide a direct route to higher fidelity by enabling faster entangling operations relative to the single-body coherence time. Placing molecules in an optical lattice would allow spacings below the diffraction limit of our objective while reducing site-to-site positional fluctuations.

\textit{Magic-wavelength trapping.} Operating the array at a magic wavelength, where the differential AC Stark shift between the qubit states vanishes, would suppress light-shift-induced dephasing and extend single-particle coherence time toward the black-body radiation (BBR) limit~\cite{ruttley2025long}.

\textit{Black-body radiation correction.} In the present experiment, single-body coherence is partially limited by black-body-induced excitation between vibrational states, with a room-temperature $1/e$ lifetime of $\tau_{\mathrm{BBR}} \approx 3.4$~s. Of the total experimental sequence, the dipolar gate occupies only $\sim 12$~ms, while the remaining overhead from shelving, tweezer rearrangement, and state preparation amounts to $\sim 100$~ms, contributing a population-loss floor to the measured fidelity. One route to correct this contribution is at the state-preparation and measurement (SPAM) level by shelving $\upstate$ into a long-lived rotational manifold (e.g. $\lvert{N=0, F=1, m_F=1}\rangle$) and resonantly clearing thermally repopulated vibrationally excited states before readout~\cite{holland2024demonstration}. A cryogenic enclosure~\cite{schymik2021single, zhang2025high} or an appropriately designed conducting enclosure~\cite{wu2023millisecond} could further suppress the relevant black-body photon density and extend $\tau_{\mathrm{BBR}}$ by orders of magnitude, removing BBR as a relevant limit on the timescales considered here.


\end{document}  

\end{document}